Complete title:

# Real-time stain-free classification of cancer cells and blood cells using interferometric phase microscopy and machine learning

Author names and affiliations:

Noga Nissim

Matan Dudaie

Dr. Itay Barnea

Prof. Natan T. Shaked

Department of Biomedical Engineering, Faculty of Engineering, Tel Aviv University, Tel Aviv 69978, Israel

Running headline: Cancer-blood cells stain-free classification

Corresponding author:

Natan T. Shaked

Department of Biomedical Engineering
Faculty of Engineering
Tel Aviv University
Ramat Aviv 69978
Israel

Phone/Fax: +972-3-640-7100

Email: nshaked@tau.ac.il

Research support:

Horizon 2020 European Research Council (ERC) Grant No. 678316.

# Real-time stain-free classification of cancer cells and blood cells using interferometric phase microscopy and machine learning

**Noga Nissim, Matan Dudaie, Itay Barnea, and Natan T. Shaked***

Department of Biomedical Engineering, Faculty of Engineering, Tel Aviv University, Tel Aviv 69978, Israel

*Correspondence to: Natan T. Shaked, Department of Biomedical Engineering, Faculty of Engineering, Tel Aviv University, Ramat Aviv 69978 Israel. Email: nshaked@tau.ac.il

## Abstract

We present a method for a real-time visualization and automatic processing for detection and classification of untouched cancer cells in blood during stain-free imaging flow cytometry using digital holographic microscopy and machine learning in throughput of 15 cells per second. As a preliminary model for circulating tumor cells in blood, we automatically classified primary and metastatic colon cancer cells, where the two types of cancer cells were isolated from the same individual, as well as four types of blood cells. We used low-coherence off-axis interferometric phase microscopy and a microfluidic channel to quantitatively image cells during flow. The acquired images were processed and classified based on their morphology and quantitative phase features during the cell flow. We achieved high accuracy of 92.56% for distinguishing between the cells, paving the way for future automatic label-free cancer cell classification in blood samples.

## Key terms

Digital holographic microscopy; Quantitative phase microscopy; Machine learning; Cell classification; Imaging flow cytometry; Liquid biopsy; Cancer cells; Blood cells.

## Introduction

The identification of circulating tumor cells (CTCs) in liquid biopsies has major prospective importance in diagnostic assessments and personalized therapeutic treatments of cancer. CTCs are highly specialized cells that may appear in small numbers in the blood stream, and originate from both primary and metastatic lesions (1,2). Therefore, CTCs can be potentially acquired from liquid biopsies, such as blood tests taken in simple routine lab procedures (3). Characterization of these rare, disease-associated cells can significantly contribute to cancer detection and the evaluation of cancer progression, as well as provide clinical information on the chosen therapy effectiveness (4). Most approaches dealing with the identification of specific cell types rely on using unique antigens or contrast agents. Specifically, in fluorescence-activated cell sorting (FACS), cells are labeled with fluorescent markers for obtaining molecular specificity (5). Attachment of fluorescent markers to antibodies that recognize target features in the cell is necessary for the unequivocal identification. The problem is that certain cell types lack these essential antibodies (6). Furthermore, the attachment of markers to the cell membrane can cause unwanted chemical interactions that might change the cell characteristics and, as a result, damage the validity of the measurement (7).

Stain-free measurement methods for the identification of different cell types overcome these problems by enabling noninvasive measurements of the cell based on the cell intrinsic properties, without using exogenous contrast agents. The refractive index (RI) of the cell is an intrinsic optical parameter that describes how fast light travels through the cell. The RI is correlated with other cell biophysical properties, including mechanical and electrical cell parameters. It represents the intracellular dry mass and concentration of the cell, and also provides valuable information about the inherent morphological organization. Various optical techniques can measure the cell RI noninvasively. One of these methods is interferometric phase microscopy (IPM). IPM, also called digital holographic microscopy and quantitative phase microscopy, measures the quantitative phase profile of the cell. This profile is proportional to the optical path delay (OPD) profile of the cell, which is equal to the product of the cell physical thickness and the difference between the integral refractive indices of the cell and the surrounding media. By acquiring the OPD profile, IPM enables visualizing cells and part of the inner cell organelles without the use of exogenous contrast agents, such as fluorescent markers, as well as classification of cells (8–12).

Although conventional flow cytometry (13) provides an extremely high throughput of up to 100,000 cells per second, it typically only provides a single value per fluorescence marker per cell (14). Imaging flow cytometry (IFC) has become a resurgence of interest due to its high-throughput and multi-parametric analysis at the single-cell level, which is based on the fact that image of the cell is captured during its flow. Typically, IFC uses exogenous contrast agents as well for morphological cell evaluation. However, in the past few years, there have been many advances in the development of label-free IFC for the analysis of cellular populations based on individual cell images. This includes the analysis of cancer cells (15–17), blood cells (18–22), cell cycle (23), cell differentiation (24) and drug response (25).

The combination of holography with IFC provides a label-free imaging technique, for cell analysis and classification during cell flow (26). Some studies used holographic IFC to measure the characteristics of different types of cancerous cells; Min *et al.* integrated a digital holographic microscopy system with a conventional flow cytometer to analyze two types of pancreatic tumor cells (27). Lee *et al.* reported high-throughput IFC by combining a quantitative phase imaging platform with time-stretch optical microscopy for the classification of human leukemic cell types (28). In addition, Zhao *et al.* presented an optical microfluidic cytometry scheme for label-free detection of cells, which is based on self-mixing interferometry technique (29). Merola *et al.* demonstrated that by exploiting the rolling of cells while they flow along a microfluidic channel, it is possible to obtain single-cell interferometric tomography for red blood cells (30). Various studies combined label-free IFC with machine-learning algorithms. Li *et al.* developed a lens-free flow cytometer based on holography for analysis of single leukocytes (31). Ugele *et al.* reported a holographic flow cytometry method for label-free differentiation of leukocytes (32). In particular, deep learning was used for classification; Göröcs *et al.* investigated objects inside a continuously flowing water sample by holographic flow cytometer (33). Chen *et al.* showed that high-throughput label-free classification of T-cells (one type of white blood cells) against colon cancer cells can be achieved through a combination of time-stretch microscopy and deep learning (34).

In this paper, we describe a real-time stain-free classification method of different types of cancerous cells from different types of blood cells using label-free holographic flow cytometry. Most importantly, the proposed technique includes rapid automated cell processing during cell visualization and flow, with high discriminative power on the level of the individual cell. To evaluate and validate the suggested system, we conducted experiments using human blood cells mixed with colorectal cancer cells. For cancer cells, we have chosen pairs of cell lines taken from the same individual, to avoid differences that are related to changes between people's organs and disease expression. The custom-built optical system can acquire single-cell holograms during flow and analyze them in real-time by applying image processing and machine learning. For training the classifiers, we acquired in advance off-axis holograms of each cell type separately using IPM and extracted the associated OPD profiles. Then, we have created a database for training a machine-learning classifier based on support vector machine (SVM) and found features that differentiated cancerous cells from a heterogeneous blood sample. This SVM model was then used for real-time classification of heterogeneous population of cells during flow.

## Materials and Methods

Optical system for quantitative phase imaging of flowing cells

We built an IPM system to rapidly capture the OPD maps of primary and metastatic cancer cells, as well as different types of blood cells *in vitro*. As shown in Fig. 1, we used a low-coherence Mach-Zehnder imaging interferometer with an off-axis configuration. Using this system, it is possible to reconstruct the sample complex wavefront from a single camera exposure, and thus it is suitable for the acquisition of rapid dynamics, such as cells during flow. The optical system is illuminated by a low-coherence light source, consisted of a supercontinuum laser source (Fianium SC-400-4), coupled to an acousto-optic tunable filter (AOTF), aligned to emit 650 nm with a spectral bandwidth of 7nm. The beam is split into a reference and a sample beams by BS1. The sample beam passes through the sample S, containing cells flowing in a microfluidic channel (ibidi, μ-Slide VI 0.4), as shown in Fig. 1 inset. The red dashed rectangle illustrates the field of view (FOV) imaged onto the camera. This beam is then magnified by microscope objective MO (Newport, 40×, 0.85 NA). In parallel, the reference beam propagates through identical microscope objective MO. Both beams are projected through tube lens TL (f=200 mm) onto a CMOS digital camera (Thorlabs, DCC1545M). Since low-coherence illumination is used to minimize spatial noise and parasitic interferences, retroreflectors RRs on micrometers are utilized to adjust the beam paths of the sample and reference beams, allowing us to obtain off-axis interference on the camera. These two beams interfere on the camera at a small off-axis angle and induce straight off-axis fringes. In addition, the chosen spectral bandwidth allowed us obtaining acceptable interference fringe modulation on the entire camera sensor.

Cell culture and sample preparation

We imaged and analyzed two types of cancer cells and four types of blood cells. For cancer cells, we used a pair of isogenic cancer cell lines: colorectal adenocarcinoma colon cells, SW-480 (ATCC CCL-228), and metastatic from the lymph node of colorectal adenocarcinoma cells, SW-620 (ATCC CCL-227). This cell line was chosen for the comparison of primary cancer cells versus metastatic cells since they both originated from the same person. The growth medium used for the cancer cell pair is Dulbecco's Modified Eagle's Medium (DMEM) (ATCC, SN. 30-2002) supplemented with 10% fetal bovine serum (FBS) (BI, SN. 04-007-1A). The cell lines were first incubated under standard cell culture conditions at 37°C and 5% $CO_2$ in a humidified incubator, until 80% confluence was achieved.

The Tel Aviv University's IRB approval for ordering blood samples was obtained. Blood samples from the blood bank were used to isolate four types of blood cells: erythrocytes, lymphocytes, monocytes, and granulocytes. We prepared a dilution medium composed of phosphate-buffered saline (PBS) (BI, SN. 02-023-1A), supplemented with 2mM ethylenediaminetetraacetic acid (EDTA) (BI, SN. 01-862-1B), 10% FBS and 6% nycodenz (Alere Technologies AS, SN. 1002424). EDTA prevents bacteria growth inside the sample; FBS keeps the vitality of the cells for a longer time; Nycodenz is added to compare the density of the dilution medium to the cell density, to achieve uniform distribution of cells inside the sample while flowing in the microfluidic channel. Since erythrocytes population in the blood is very high (~1000:1), we dissolved the blood in dilution medium with a proportion of 1:250. The rest of the blood cells were isolated by the following protocol. We centrifuged 2ml of blood, 2ml PBS supplemented with 2mM EDTA and 2ml of Ficoll-Paque PLUS (GE Healthcare, SN.17-1440-02) for 20 minutes. Then, we extracted two layers: (1) Middle layer (buffy coat), containing most of the white blood cells and platelets, for the isolation of lymphocytes and monocytes. (2) Lower layer, containing granulocytes and erythrocytes.

The buffy coat was centrifuged with PBS supplemented with 2mM EDTA, and at the end of the process, the supernatant layer was removed. We repeated the last stage in order to discard platelets and remained with white blood cells only. For the isolation of lymphocytes and monocytes, we used commercial kits (lymphocytes CD4+: EasySep, SN. 17952, monocytes: EasySep, SN. 19359), according to the manufacturer's instructions.

The lower layer was also centrifuged with PBS supplemented with 2mM EDTA twice, while removing the supernatant layer every time. Then, ammonium-chloride solution (STEMCELL, SN. 07850) was added, and the whole solution was incubated on ice for one hour in order to achieve lysis of erythrocytes. Afterward, the solution was washed and centrifuged again with PBS supplemented with 2mM EDTA for the isolation of granulocytes.

Each of the homogeneous samples was dissolved with a dilution medium according to the desired concentration in the microfluidic channel.

<u>Data analysis</u>

To extract the quantitative phase maps from the acquired off-axis image holograms, we used the off-axis interferometry Fourier-based algorithm (35), which included a digital 2-D Fourier transform, filtering one of the cross-correlation terms, and an inverse 2-D Fourier transform, where the argument of the resulting complex-wavefront matrix was the wrapped phase of the sample. To compensate for stationary aberrations and field curvatures, we subtracted from the wrapped phase map of the sample a phase map that is extracted from a hologram acquired with no sample present. We then applied the unweighted least-squares phase unwrapping algorithm to resolve the $2\pi$ phase ambiguities (36). The resulting unwrapped phase map was multiplied by the wavelength and divided by $2\pi$ to obtain the quantitative OPD map of the sample. This OPD map is defined as follows:

$$(1) \quad OPD_c(x, y) = [\bar{n}_c(x, y) - n_m] \times h_c(x, y),$$

where $n_m$ is the RI of the medium, $h_c$ is the thickness profile of the cell, and $\bar{n}_c$ is the cell integral RI, which is defined as follows:

$$(2) \quad \bar{n}_c(x, y) = \frac{1}{h_c} \int_0^{h_c} n_c(x, y, z) dz.$$

In the resulting OPD profile, the cell area was isolated by a simple threshold, followed by a morphological dilation. In cases of frames with no cell, the classification process was not needed. Therefore, we applied another threshold for the minimum size of the connected component. We also applied a maximum size threshold in cases of attached cells that could not be classified as one object. Partial images of cells on the edges of the FOV were not classified as well. Using the above described methods, we created a dataset containing the OPD information across the cell areas only, and calculated the different parameters that were based directly on the OPD map defined in Eq. (1), without decoupling the cellular thickness profile from the refractive index as a prior stage.

The features that have been extracted from each OPD map divide into two categories: (1) 2D morphological features; and (2) 3D quantitative features. The 2D morphological features are based on the binary image indicating the cell area only. The 3D quantitative features are based on the OPD map across the cell area. These features are presented in Table 1. The 3D quantitative features rely on previous works that demonstrate the ability to distinguish between the different stages of the cell lifecycle, as well as other biological phenomena (11,37,38).

<u>Machine-learning algorithms</u>

For classification, the more complex task is classification of white blood cells and cancer cells, since red blood cells and platelets are much easier to detect, since they are very different than cancer cells. We have created a data set of ~4,000 OPD maps of two types of colorectal cancer cells (SW480 and SW620) and four types of blood cells (granulocytes, lymphocytes, monocytes, and erythrocytes). Some examples of OPD maps from the dataset are shown in Fig. 2.

The classifier has been built using SVM, which is a common machine-learning classification algorithm that is based on features extraction. The goal of the SVM algorithm is to find a hyperplane in the features space that distinctly classifies the data points (39,40).

We have built two types of SVM classifiers and compared their performance. The first classifier is a simple multi-class SVM that classifies all six types of cells in parallel, which is coined as 'one-step SVM'. The second classifier is combined of three sub-classifiers: (1) SVM 1: classifying between cancer and blood cell; (2) SVM 2: classifying between two types of cancer cells; and (3) SVM 3: classifying between four types of blood cells. We coined the second classifier as 'two-step SVM', because its classification process is carried out in two stages: SVM 1 receives the OPD map of the cell and categorizes it in a general manner as 'cancer' or 'blood' cell. Afterward, a more specific classification is performed by SVMs 2 and 3, which receive their inputs according to SVM 1 result. Schemes of the one-step SVM and the two-step SVM are described in Fig. 3. All models were trained and tested on 80% and 20% of the dataset, respectively. The training was done using the10-fold cross-validation procedure.

We also used principal component analysis (PCA), which is a common method for dimension reduction and for finding highly discriminative features. The PCA method is based on projecting the data onto a lower-dimension subspace, and receiving new features, which are linear combinations of the original features. The first principal component has the largest possible variance of the data, and therefore enables better discrimination between the classes, the second principal component has the second largest variance of the data, and so on (41).

The whole method of classification (training and real-time testing) described above was implemented using MATLAB on a standard CPU.

## Results

We used the low-coherence IPM system presented in Fig. 1 to quantitatively image all six cell types during flow in a throughput of 15 cells per second, acquired their OPD maps and classified them in real-time, using a standard personal-computer CPU on a MATLAB platform. The experiments work were divided into two main parts: first, we flowed homogeneous samples of each cell type to create a supervised dataset for training and testing the models; second, we flowed different heterogeneous samples, containing several cell types together, to validate the classifier performance in real-time.

Prior to the analysis of the OPD maps, we applied the segmentation image-processing procedure, described in the Methods section, to track the cell area during flow. Next, we extracted the twenty features, mentioned in Table 1, from the cell OPD area selected by the segmentation process. During the training process, the twenty hand-crafted features were used as an input for PCA analysis in order to extract the best combinations of these features, which were the most useful for classification between various cell types. The best classification results were obtained for the eight, six, ten and thirteen first principal components for SVM 1, SVM 2, SVM 3 and one-step SVM, respectively. Figure 4 demonstrates the separation in the 2D PCA space for all four SVM models. The $x$ and $y$ axes represent the two first principal components, which have most of the variance of the data. In practice, the separation between the cell types was even better than shown in this illustration, since more than two principal components were practically used for classification, as elaborated above. The new highly-discriminative features provided by PCA were then used as an input for the SVM models.

For comparison, we trained all SVM models separately based the: (1) 2D morphological features; and (2) 2D morphological and 3D quantitative features together. Table 2 presents the accuracies of these two assays for all SVM models. As can be seen, the accuracy is higher when considering both the 2D morphological features and the 3D quantitative features for all trained models, demonstrating the advantage of using quantitative phase images for classification rather than simple 2D imaging. Table 3 presents the precision of performing wrong and right classifications with all SVM models combined with PCA on the test set, taking into account both the 2D morphological features and the 3D quantitative features. As seen in Table 2 and Table 3, the two-step classifier exhibits the best overall accuracy when examining it on the test set.

Next, we compared the performance of the one-step and the two-step classifiers on different samples during flow. The ability to analyze stain-free isolated cells is important for flow-cytometry via quantitative imaging of cells during flow. Figure 5 presents snapshots of OPD maps taken while the cells are flowing in the microfluidic channel, with the real-time quantitative-imaging visualization and classification results during cell flow. See Supplementary Video 1 for several periods of flowing of homogeneous samples of SW480 and granulocytes. The classification result and the confidence score of the classifier prediction are presented on the top. A green caption appears if the cell has been classified as a cancer cell, and a red caption appears if to the cell has been classified as a blood cell.

Next, we mixed an even amount of SW480 and SW620 cells and made them flow in the channel. See Supplementary Video 2 and Supplementary Video 3 for several periods of flowing of these heterogonous samples. Table 4 presents the classification results of three classifiers: one-step SVM, two-step SVM and SVM 2. As expected, SVM 2 achieved the best results for classifying between the two cancer classes only. Observing the classification results of the one-step SVM and the two-step SVM classifiers, both classified SW480 cells close to 50%, the correct mixture percentage, while most of the wrong predictions occurred when misclassifying the SW620 cells as granulocytes. This misclassification can be explained by the fact that granulocytes are relatively large cells and characterized with high-variability of OPD values, similarly to the SW620 cells. Between the two main classifiers, the two-step SVM reached better results.

Next, we used a homogeneous sample of granulocytes and imaged them during flow. Table 5 shows the classification results of three classifiers: one-step SVM, two-step SVM and SVM 3. Unsurprisingly,

SVM 3 achieved the best results, since it classified between blood cells only. Here as well, the two-step SVM achieved better results than the one-step SVM.

The average processing times for each step in the algorithm were as follows: (1) 0.028 sec for the reconstruction of the unwrapped phase profile, (2) 0.025 sec for cell segmentation and features extraction, (3) 0.01 sec and 0.095 sec for cell classification by the two-step SVM and the one-step SVM, respectively. Although the two-step classifier included two SVM models, while the one-step classifier included only one, the two-step SVM exhibited faster execution time. Combining all, the total processing times for each off-axis hologram containing 1 megapixel is 0.063 sec and 0.148 sec for the two-step SVM and the one-step SVM, respectively.

These results demonstrate the ability of the presented automatic algorithm to classify cancer in different cancer stages and white blood cells using flow-cytometry combined with machine learning, using OPD-map-based features. We achieved high classification rates for stain-free cells during real-time flow. The accuracy values and prediction precisions correspond with the separation between the groups presented in the 2D PCA space (see Fig. 4, Tables 2-3), where the two-step SVM classifier manifested better classification results than the one-step classifier.

## Discussion and conclusion

Early detection of cancer greatly increases the chances of successful treatment and survival. Most diagnostic techniques based on liquid biopsies rely on analyzing the cells after antibodies attachment to specific tumor antigens, which might affect the cell behavior. Our work can distinguish cancer cells from blood cells, and primary cancer cells from metastatic cancer cells, without using external labeling during cell flow.

Our approach is based on stain-free quantitative phase microscopy for the identification and classification of live and unattached cancerous cells inside a blood sample. We outlined the diagnostic framework and the detailed instrumentation layout. For cell classification, we have employed machine-learning techniques that included PCA-based cell type representation followed by an SVM-based classification.

We identified OPD-based features of four types of blood cells and two types of colorectal cancer cells. The cancer cells in our study were taken from the same individual. We showed the feasibility to distinguish between these cells based on 2D morphological features and 3D quantitative features, demonstrating that using the 3D features, better classification results are obtained.

We created two types of classification models, one-step SVM and two-step SVM, and compared between their performances. The two-step SVM exhibited a bit higher accuracy on the test set and also achieved better results in the real-time flowing samples. This result can be explained by the fact that SW620 cells and granulocytes are likely to be more similar than SW620 to SW480 and granulocytes to the rest of the blood cells. Because the one-step SVM is a multi-class SVM that classifies all six types of cells in parallel, its misclassification usually occurs between SW620 cells and granulocyte, while the two-step SVM avoids this task by splitting the classification into two parts.

Comparing the execution times between the two classifiers, the classification process in the two-step SVM was more than nine times faster than in the one-step SVM, although the two-step SVM included two classification steps while the one-step SVM included only one. This difference can be explained by the fact that each classification step in the two-step SVM has fewer classes than in the one-step SVM. This result agrees with a common approach that hierarchical SVMs can achieve significantly better and more stable performance than other strategies, as well as high computational efficiency (42).

By automating the processes of image acquisition and cell identification, our approach enables higher system throughputs than conventional methods. In this work, we presented a proof-of-concept throughput of 15 cells per second. We suggest some instrumentation layout options for future-throughput-improvements in three levels. First level will be using better quality computational instruments. The presented approach was implemented on a standard personal computer CPU. Applying our machine-learning algorithm with a suitable graphic-processing-unit (GPU) or a more powerful CPU will improve execution time and achieve better throughput. The second level is using faster sensing methods. For example, serial time-encoded amplified imaging was proved to be an indispensable tool for studying dynamical events such as chemical dynamics in living cells and microfluidics. This imaging method overcomes limitations of conventional CCDs and offers framerates that are more than 1,000 times faster (43). The third level for speeding up processing time is using holographic compression. In the past years, we presented methods that can speed up holographic acquisition times by multiplexes two orthogonal off-axis holograms on the camera at once, where the digital reconstruction, including spatial filtering and two-dimensional phase unwrapping on a decreased number of pixels, can be performed on both holograms together, without redundant operations (44,45). We have lately shown that up to six holographic channels can be multiplexed without spatial-frequency overlap, providing the possibility to acquire off-axis holograms up to six times faster with the same number of camera pixels (46). In the future, the demonstrated platform can help in early detection of cancer and in identifying different stages of oncogenesis. Detecting CTCs in blood is very challenging, since it requires a much higher acquisition and processing throughput (47). Other types or liquid biopsies are more likely to be useful first, such as detecting CTCs in urine sample for bladder cancer patients.

To conclude, we presented a stain-free imaging flow cytometry-based method with automatic real-time classification of untouched cancer cells in blood by using digital holographic microscopy and machine learning. The proposed method is noninvasive, automatic, and provides high-discriminative power on the level of the individual cell. It is implemented in real-time and with applicable modifications may achieve higher-throughput. In the future, we believe that the demonstrated platform can help in detection of cancer and to identify different stages of oncogenesis.

## Acknowledgments

This study was supported by Horizon 2020 European Research Council (ERC) grant no 678316.

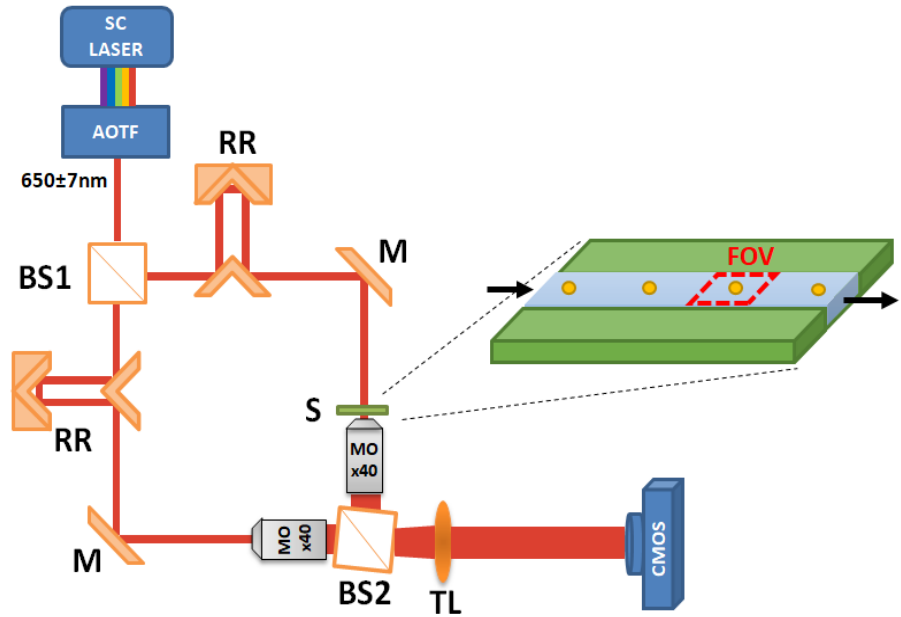

**Figure 1.** The optical system used for quantitative phase imaging of flowing cells.SC, Supercontinuum source. AOTF, Acousto-optic tunable filter. BS1, BS2, Beam splitters. M1, M2, Mirrors. RR, Retroreflector. S, Sample. MO, Microscope objective. TL, Tube lens. CMOS, Digital camera. FOV, Field of view.

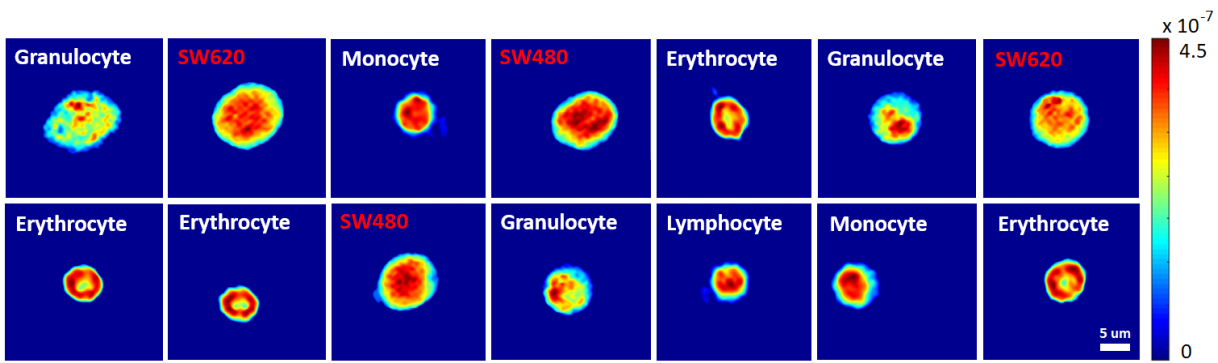

**Figure 2.** Examples from the acquired database of quantitative phase images of cancer (red caption) and blood (white caption) cells. Colorbar represents OPD values in meter.

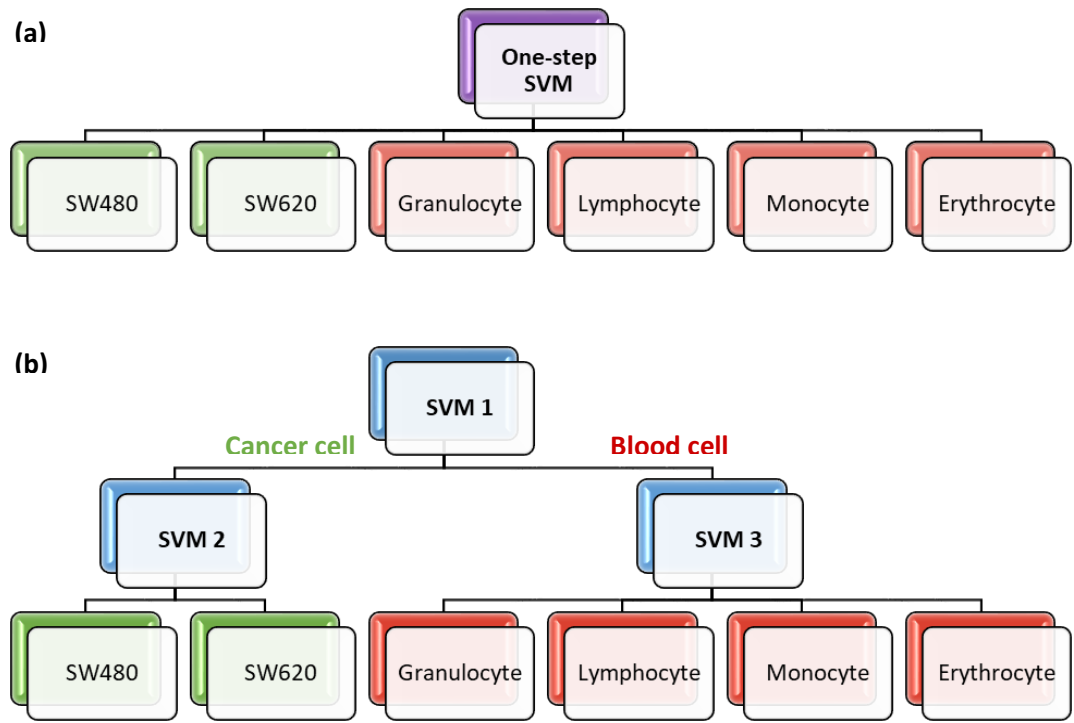

**Figure 3.** Block diagram schemes of two types of classification processes. (a) One-step SVM classifier. (b) Two-step SVM classifier.

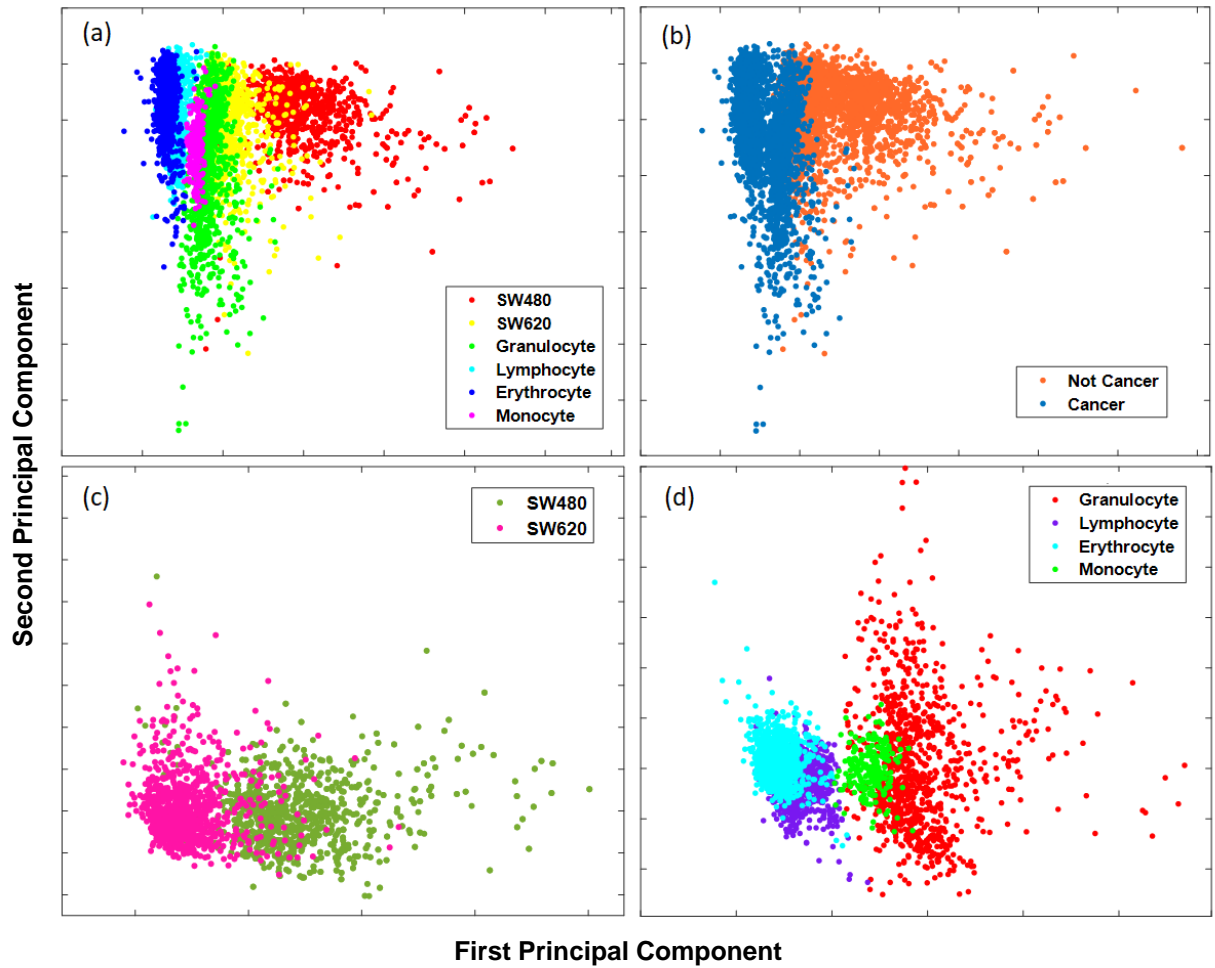

**Figure 4.** Scatter spot in the 2D PCA space for all four classifiers, where the *x* and *y* axes represent the two first principal components. (a) One-step SVM. (b) SVM 1. (c) SVM 2. (d) SVM 3. Note that the actual separations between the groups are even better than presented, since only the first two principal components are used in this illustration, but the actual classifications have used more principle components.

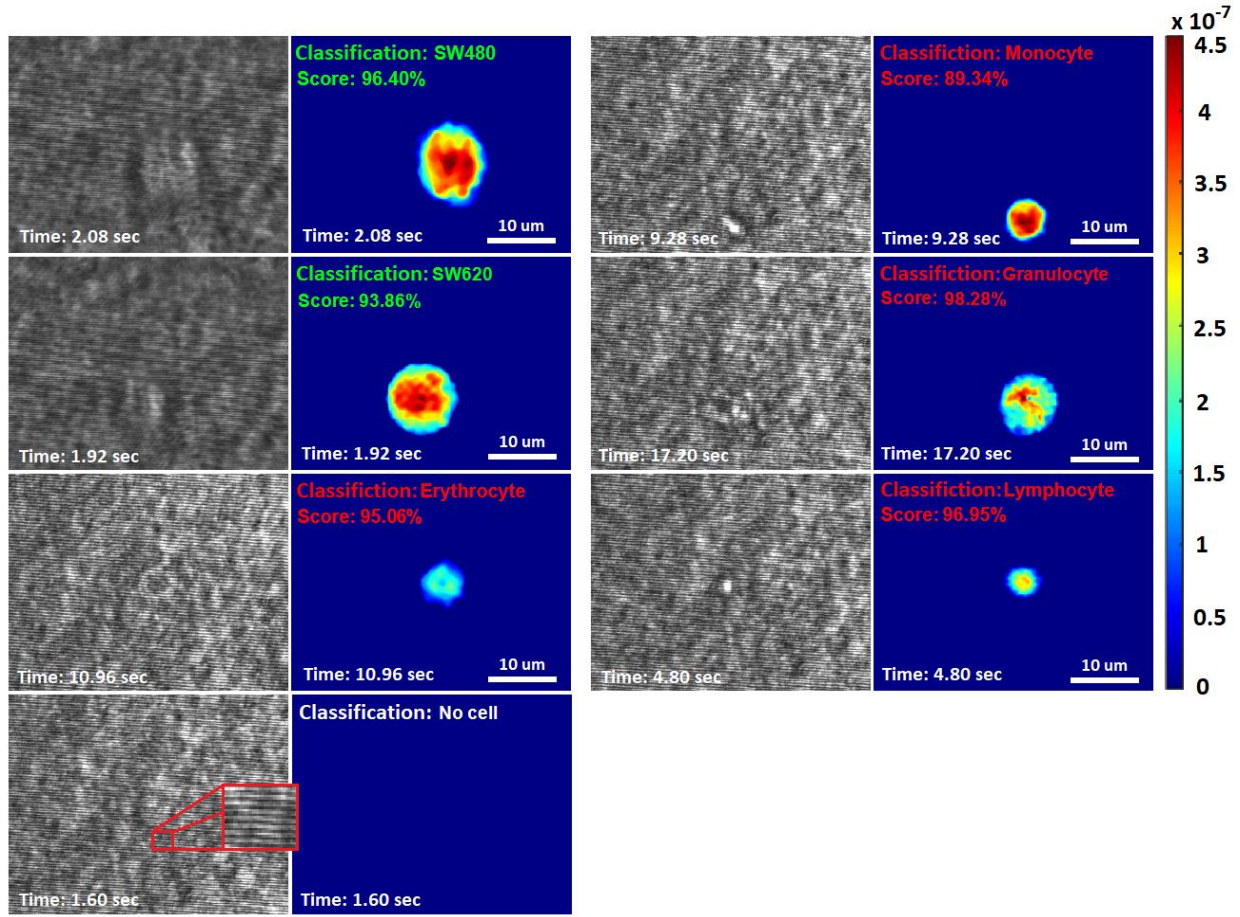

**Figure 5.** Snapshot from the real-time classification videos, presenting quantitative phase microscopy of live unlabeled cells flowing in a microfluidic channel. Each pair of images depicts the same frame, where the left image is the original off-axis hologram captured by the camera, and the right image is the OPD profile processed in real-time. The classification results are written on the top. A green caption refers to cancer cell prediction, red caption refers to blood cell prediction, and white caption refers to frames with no cell. Colorbar represents OPD values in meter.

| | Cell 2D morphological features | Cell 3D quantitative features |
|---|---|---|
| 1 | Diameter | Mean |
| 2 | Area | Energy |
| 3 | Major axis length | Volume |
| 4 | Minor axis length | Area to Volume ratio |
| 5 | Minor to Major ratio | Dry mass |
| 6 | Convex Area | Variance |
| 7 | Eccentricity | Kurtosis |
| 8 | Circularity | Skewness |
| 9 | | Contrast |
| 10 | | Entropy |
| 11 | | Homogeneity |
| 12 | | Correlation |

**Table 1.** 2D (left) and 3D (right) handcrafted features extracted in real time from the OPD profile of the flowing cells.

| Model | Total accuracy | |
|---|---|---|
| | 2D morphological features | 2D morphological + 3D features |
| SVM 1 | 92.88 % | 98.22% ↑ **5.34%** |
| SVM 2 | 95.6 % | 96.98% ↑ **1.38%** |
| SVM 3 | 74.68 % | 91.77 % ↑ **17.09%** |
| One-step SVM | 78.46 % | 91.3 % ↑ **12.84%** |
| Two-step SVM | 77.6 % | 92.56% ↑ **14.96%** |

**Table 2.** Accuracy results when using the 2D features and the 3D features for all SVM models. The improvement obtained for each classification method when also using the 3D features in comparison to using the 2D features only is indicated in green.

| Model | Class | Right prediction | Wrong prediction |
|---|---|---|---|
| SVM 1 | Cancer | 97.3% | 2.7% |
| | Not Cancer | 99.1% | 0.9% |
| SVM 2 | SW480 | 97.5% | 2.5% |
| | SW620 | 96.3% | 3.7% |
| SVM 3 | Granulocyte | 90% | 10% |
| | Lymphocyte | 97.8% | 2.2% |
| | Erythrocyte | 91.2% | 8.8% |
| | Monocyte | 82.6% | 17.4% |
| One-step SVM | SW480 | 97.8% | 2.2% |
| | SW620 | 88.6% | 11.4% |
| | Granulocyte | 88.8% | 11.2% |
| | Lymphocyte | 98.6% | 1.4% |
| | Erythrocyte | 86.2% | 13.8% |
| | Monocyte | 82.6% | 17.4% |
| Two-step SVM | SW480 | 97.5% | 2.5% |
| | SW620 | 90.1% | 9.9% |
| | Granulocyte | 87.6% | 12.4% |
| | Lymphocyte | 97.8% | 2.2% |
| | Erythrocyte | 91.2% | 8.8% |
| | Monocyte | 82.6% | 17.4% |

**Table 3.** Precision of wrong and right predictions for all SVM models on the test set.

| Class | One-step SVM | Two-steps SVM | SVM 2 |
|---|---|---|---|
| SW480 | 42.21 % | 46.5 % | 50.26 % |
| SW620 | 10.3 % | 29 % | 49.74 % |
| Granulocyte | 46 % | 21.75 % | 0 % |
| Lymphocyte | 1 % | 2 % | 0 % |
| Erythrocyte | 0.5 % | 0.5 % | 0 % |
| Monocyte | 0 % | 0.25 % | 0 % |
| **Total cancer cells** | **52.5 %** | **75.5 %** | **100 %** |
| **Total blood cells** | **47.5 %** | **24.5 %** | **0 %** |

**Table 4.** Classification results of a sample of containing a 1:1 mixture of flowing SW480 and SW620 cells.

| Class | One-step SVM | Two-steps SVM | SVM 3 |
|---|---|---|---|
| SW480 | 0.59 % | 1.77 % | 0 % |
| SW620 | 4.12 % | 1.64 % | 0 % |
| Granulocyte | 78.34 % | 80.12 % | 83.04 % |
| Lymphocyte | 4.5 % | 3.29 % | 4.5 % |
| Erythrocyte | 0.2 % | 0.3 % | 0.2 % |
| Monocyte | 12.25 % | 12.88 % | 12.25 % |
| **Total cancer** | **4.71 %** | **3.41 %** | **0 %** |
| **Total blood** | **95.29 %** | **96.59 %** | **100 %** |

**Table 5.** Classification results of a homogeneous sample of flowing granulocytes.